\documentclass[aps,twocolumn,prx,amsmath,amssymb,notitlepage,superscriptaddress,longbibliography,nofootinbib,longbibliography]{revtex4-2}

\usepackage{graphicx}
\usepackage{dcolumn} 
\usepackage{bm} 
\usepackage{hyperref}
\usepackage[mathlines]{lineno}

\usepackage{xcolor}
\definecolor{darkblue}{rgb}{0,0,0.6}
\definecolor{darkred}{rgb}{0.6,0,0}
\definecolor{darkgreen}{rgb}{0,0.6,0}
\definecolor{bleu}{rgb}{0,0.44,0.72}
\usepackage{setspace}
\usepackage{ulem}
\usepackage{enumerate}

\newcommand{\cD}{\mathcal{D}}

\newcommand{\cG}{\mathcal{G}}

\newcommand{\bff}{\textbf{f}}

\newcommand{\bu}{\textbf{u}}

\newcommand{\brho}{\bar{\rho}}

\newcommand\rmd{\mathrm{d}}

\newcommand\bfsigma{\boldsymbol{\sigma}}

\newcommand\bfJ{{\bf J}}

\newcommand\bfr{{\bf r}}

\newcommand\bfu{{\bf u}}

\usepackage{tikz}

\definecolor{user_yellow}{HTML}{FFBE0B}

\begin{document}

\title{Synthetic quorum sensing and absorbing phase transitions in  colloidal active matter}

\author{Thibault Lefranc}
\altaffiliation[These authors contributed equally to this work.]{}
\affiliation{ENSL, CNRS, Laboratoire de physique, Universit\'e de Lyon, F-69342 Lyon, France}
\author{Alberto Dinelli}
\altaffiliation[These authors contributed equally to this work.]{}
\affiliation{Department of Biochemistry, University of Geneva, 1211 Geneva, Switzerland}
\affiliation{Université Paris Cité, MSC, UMR 7057 CNRS, 75013 Paris, France}
\author{Carla Fernandez Rico}
\affiliation{Department of Materials, ETH Z\"urich, 8093 Z\"urich, Switzerland}
\author{Roel P. A. Dullens}
\affiliation{Institute for Molecules and Materials, Radboud University Heyendaalseweg 135, 6525 AJ Nijmegen, The Netherlands}
\author{Julien Tailleur}
\altaffiliation[Corresponding authors.]{ Email: {denis.bartolo@ens-lyon.fr and jgt@mit.edu}}
\affiliation{Department of Physics, Massachusetts Institute of Technology, Cambridge, Massachusetts 02139, USA}
\author{Denis Bartolo}
\altaffiliation[Corresponding authors.]{ Email: {denis.bartolo@ens-lyon.fr and jgt@mit.edu}}
\affiliation{ENSL, CNRS, Laboratoire de physique, Universit\'e de Lyon, F-69342 Lyon, France}

\date{\today}

\begin{abstract} 
Unlike biological active matter that constantly adapt to their environment, the motors of synthetic active particles are typically agnostic to their surroundings and merely operate at constant force.
Here, we design colloidal active rods capable of modulating their inner activity in response to crowding, thereby enforcing a primitive form of quorum sensing interactions. 
Through experiments, simulations and theory we elucidate the impact of these interactions on the phase behavior of isotropic active matter. 
We demonstrate that, when conditioned to density, motility regulation can either lead to an absorbing phase transition, where all particles freeze their dynamics, or to atypical phase separation,  where flat interfaces supporting a net pressure drop are in mechanical equilibrium. 
Fully active and fully arrested particles can then form heterogeneous patterns ruled by the competition between quorum sensing and mechanical interactions. 
Beyond the specifics of motile colloids, we expect our findings to apply broadly to adaptive active matter assembled from living or synthetic units. 
\end{abstract}

\maketitle

\pagebreak

\section{Introduction}
\noindent When in a group, microorganisms, plants,  animals and humans   regulate their dynamics in response to interactions with their neighbors.
When walking in a subway station, we adjust our pace depending on the density of the crowd that surrounds us, and organisms as elementary as \textit{E. Coli} regulate their gene expression and swimming motion in response to  chemicals released by surrounding bacteria. 
This fundamental feedback mechanism is known as quorum sensing ~\cite{miller2001quorum}.
As in all sensing process, quorum sensing relies on changes in the inner dynamics of active bodies to regulate their dynamics in response to external cues. 
Quorum sensing has spectacular consequences on the macroscopic organisation of active matter. 
In bacteria colonies, it
 results in large-scale patterns, wave propagation and biofilm formation, all determined by the interplay between bacterial activity and structural organization, see e.g.~\cite{ben2000,hammer2003quorum,daniels2004quorum,Keller2006,liu2011sequential,curatolo2020cooperative,pfreundt2023}.
By contrast, the available synthetic active particles are so rudimentary that they cannot sense their environment, the
mesoscopic structure they form  hardly ever feedback on their inner motorization and
their positions and orientations  react passively to forces and torques~\cite{howse2007,Theurkauff2012,palacci2013living,Bricard2013,Zottl2016,Zhang2017}.

While theory and simulation have shown that the interplay between structure and activity leads to robust self-organization pathways~\cite{peruani2008dynamics,cates2013active,saha2014clusters,paoluzzi2018fractal,abaurrea2018collective,o2020lamellar,paoluzzi2020information,zhang2023pulsating}, their experimental investigations have been limited to  computer-assisted control of small collections of active units~\cite{lavergne2019,Fernandez2020,Muinos2021,ben2023morphological}. 
Developing scalable active systems capable of sensing their environment and regulating  their motorization in response to structural changes remains an open challenge.

In this article, we study the phase behavior of large collections of active colloidal rods. First, we experimentally establish that colloidal rods powered by Quincke motors~\cite{Quincke1896} feature a primitive form of quorum sensing: they actively respond to density variations by   switching their internal motor on and off. 
In dilute regions, the rods roll freely on solid surfaces, whereas in crowded environments, they stand upright and arrest their motion.
We then experimentally explore  the consequences of synthetic quorum sensing by  varying independently the activity and packing fraction of our Quincke rods. 
We show that they first undergo a phase transition where an active isotropic liquid coexists with an arrested phase of  standing rods. 
Further increasing their packing fraction, despite the absence of any change in the energy injection process,  all the rods arrest their self-propulsion and form a macroscopic amorphous solid.
To elucidate our experimental findings we combine analytical theory and numerical simulations. 
We show that quorum sensing and steric interactions compete to stabilize
an atypical phase separation where flat interfaces supporting a net pressure drop are in mechanical equilibrium.
On the one hand, our primitive form of quorum sensing always induces an absorbing phase transition where  all the particles  freeze their dynamics.
On the other hand, unlike e.g. in  transitions between turbulent states ~\cite{Kazumasa2007,sano2016universal,shih2016ecological,hof2023directed}, steric interactions induce  mechanical stresses that can arrest the absorption process and allow the active and passive phases to coexist.
 
Altogether our results reveal a novel mechanism for phase coexistence in active matter that could apply to any synthetic or living system where spatial structures emerge from  sensing interactions.
\begin{figure*}
\includegraphics[width=\textwidth]{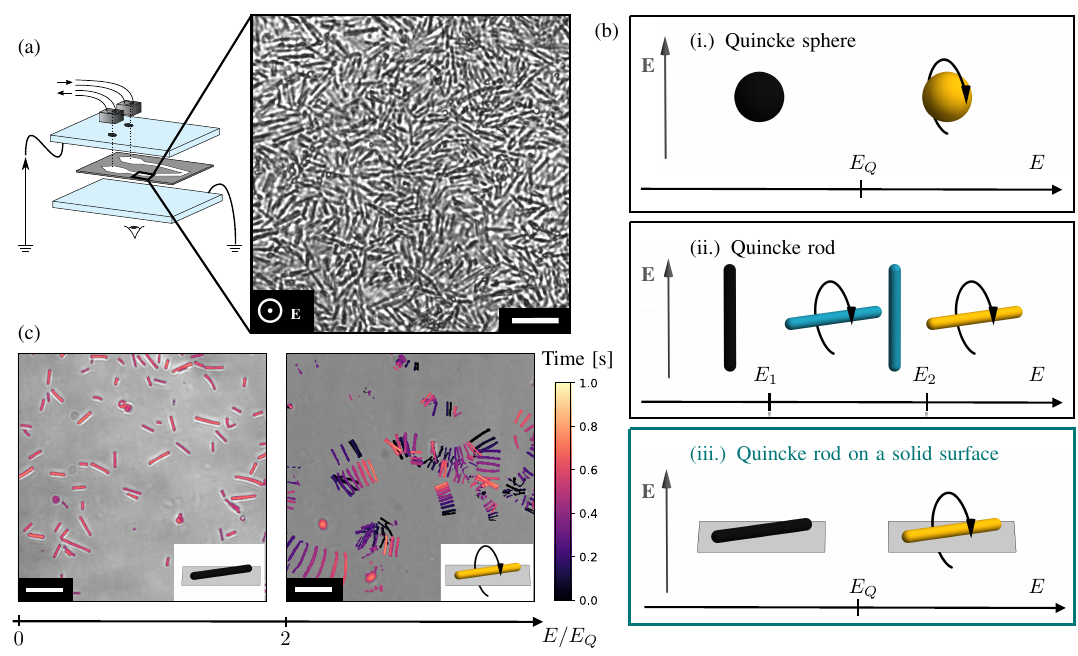}
\caption{{\bf Motorizing colloidal rods}.
  (a) (Left) Sketch of our microfluidic setup. 
  We first tilt the device  to let the rods sediment and accumulate at the tip of the V-shaped channel. 
  We then place the device in the horizontal direction and perform our observations.
  (Right) Bright Field image of the SU8 rods at density $\rho_0=0.16\,\rm \mu m^{-2}$, and voltage $V=150\,\rm V$. 
  Scale bar: $5\,\rm \mu m$.
    (b) Quincke  instabilities. 
    (i) Electrohydrodynamic instability of a dielectic sphere immersed in a conducting fluid. 
    When applying a DC  field, the static configuration is unstable when $E>E_Q$, and the sphere spins around a random axis normal to $\mathbf E$~\cite{Melcher1969}.
    (ii) When the particle is anisotropic, it points in the direction of $\mathbf E$ when $E$ is small. 
    The rod is static. 
    At large $E$, the particle aligns in a random direction normal to $\mathbf E$ and spins around its long axis. 
    At intermediate fields both states are stable and coexist~\cite{Cebers2000,Brosseau2017}
    (iii) In the vicinity of a solid surface,  rods that are dense or charged enough are driven toward the surface. 
    Their behavior mirrors that of of Quincke spheres. 
    When $E<E_Q$ the rod is static and aligned with the solid surface, above $E_Q$ it rolls along the solid wall in the direction of its short axis 
    (c) Superimposed images of SU-8 colloidal rods (length: $3.7\,\rm \mu m$ diameter: $0.55\,\rm \mu m$) sedimented on a solid electrode for $E/E_{Q}=0,\, 2$. 
    The $\mathbf E$ field points in the direction normal to the images.
    The color indicates time (Time step $0.1\,\rm s$). 
    The polydispersity in the rod length  results in a distribution  Quincke thresholds. 
    We define $E_Q$ as the  field where the first rod starts rolling. Scale bar: $5\,\rm \mu m$.}
    \label{Fig1}
\end{figure*}
\section{Experiments: engineering quorum-sensing interactions with Quincke rods}
We detail the experimental methods in Appendix~\ref{app:Quincke}.
In short, to investigate the collective dynamics of Quincke rods, we synthesize 
$3.7\pm1.5\,\rm \mu m$ long photoresist cylinders of diameter $0.55\pm0.17\,\rm \mu m$ (aspect ratio: 6.7)~\cite{Fernandez2019}.
We then turn these inanimate plastic bodies into self-propelled particles using the Quincke motorization principle~\cite{Jakli2008,Bricard2013}, and observe their collective dynamics  over a range of local number density  $\rho(\mathbf r)$ in a V-shape microfluidic channel, see Fig.~\ref{Fig1}a.
Our first control parameter $\rho_0$ is the number density of colloids averaged in the 
observation window, see Fig.~\ref{Fig1}a. 
The smallest $\rho_0$ corresponds to well separated rods all lying on the bottom electrode. At the highest $\rho_0$ values the rods accumulate over two  layers and form a quasi two-dimensional material.
Our second control parameter is the magnitude of the DC voltage  which controls the rod activity ($V\in[10\,\rm V, 300\,\rm V]$). 
\subsection{Electrorotation of  non-interacting rods}
Before
dwelling into the phase behavior of interacting Quincke rods, we briefly recall their individual dynamics
previously reported in~\cite{Cebers2000,Brosseau2017}. This will prove useful to understand how these plastic cylinders can feature quorum-sensing capabilities.

Away from  solid surfaces, the dynamics of an  insulating sphere suspended in a conducting fluid is fairly simple. 
Upon the application of a DC electric field $\mathbf E$, the particle is either static when $E$ is smaller than a critical value $E_{\rm Q}$, or spontaneously spins around a random axis transverse to $\mathbf E$ when $E>E_{\rm Q}$, see Fig.~\ref{Fig1}b.
This dynamics is the result of the well-known Quincke electrorotation instability~\cite{Quincke1896,Melcher1969}.
However, when the particle are anisotropic (rods or prolate elipsoids) their dynamics is  more complex. 
There exists three regimes illustrated in Fig.~\ref{Fig1}b. 
At low $\mathbf E$ field values, the rods are static and point in the direction of $\mathbf E$. 
At strong fields, they  spin around their long axes, which point in random directions transverse to $\mathbf E$. 
At intermediate fields, both states are stable and can coexist~\cite{Cebers2000,Brosseau2017}. 
We stress that all threshold fields scale linearly with  the viscous-drag coefficient of the active spinners. 

In our experiments, the rods  lie on a horizontal solid substrate whose normal direction is aligned with the $\mathbf E$ field.
As a result, the gravitational and  electrophoretic torques acting on the rods destabilize the vertical configurations.
We thus never observe  static vertical rods in isolation, neither in colloidal rod experiments, nor using a hundred times bigger 3D printed particles, see Fig.~\ref{Fig1}c and Supplementary Movie 3.
When dilute, Quincke rods behave similarly to spherical colloidal rollers. 
When sitting on solid surfaces: they are either static and parallel to the surface at low field, or spin and roll along their short axis when $E>E_{\rm Q}$, see Fig.~\ref{Fig1}c. We stress that their direction of motion is not slaved to any external field. Isolated Quincke colloids propel autonomously along random directions.

At the single-particle level the only difference between Quincke sphere and rods is the persistence length of their rolling motion. 
In practice, the persistence length of the Quincke rods is so large that we cannot measure it accurately in our devices~\cite{Bricard2013}. 
We could  therefore naively expect the collective behavior of our active rods to mirror that of spherical  rollers  that realize a prototypical example of a flocking transition~\cite{Bricard2013}. 
Our experimental observations proved to be more surprising. 

\subsection{Phase behavior of interacting Quincke rods}
\begin{figure*}
\includegraphics[width=\textwidth]{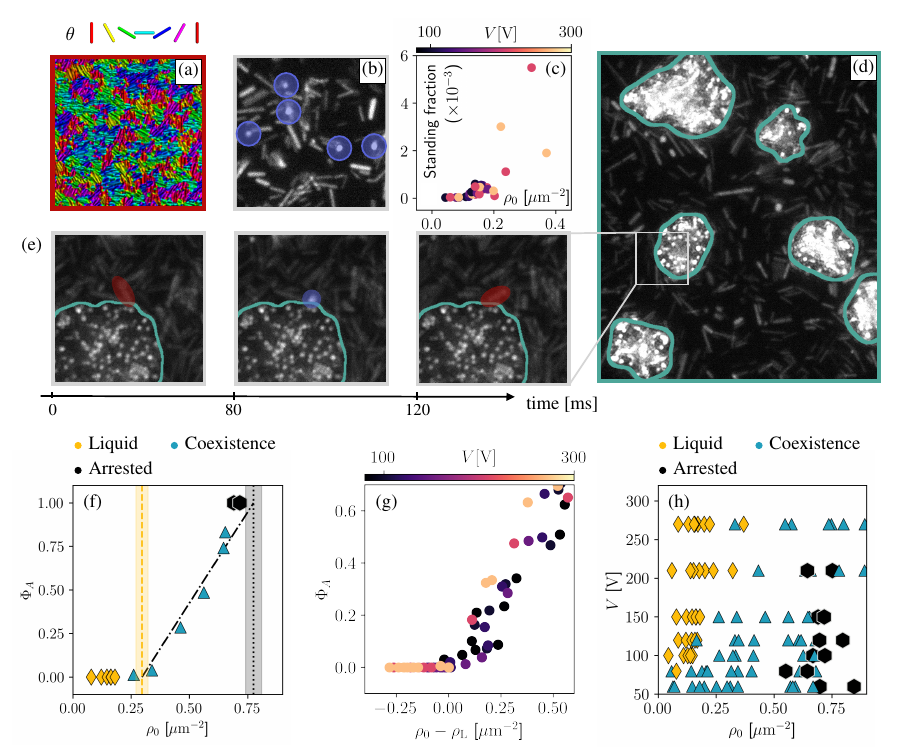}
\caption{{\bf Phase behavior of interacting Quincke rods}.
    (a) Bright field picture of a homogeneous liquid phase ($\rho_0=0.12\,\rm \mu m^{-2}$, $V=150\,\rm V$). The color indicates the local orientation of the rods measured using the method of Rezakhaniha et al.~\cite{rezakhaniha2012}.
    The active liquid is isotropic. 
    (b) Fluorescence image of the same liquid. Unlike in the gas phase, some rods have switched to a standing state: they stop spinning and point in the vertical direction (bright dots). 
    The shaded circles help locating the arrested rods. We stress that only $25\,\%$  of the rods are fluorescent, see Appendix A.
    (c) Average fraction of standing rods $\alpha/(\alpha+\beta)$ plotted against the average density $\rho_0$. 
    (d) Fluorescence image of  arrested clusters coexisting with an active liquid phase ($\rho_0=0.21\,\rm \mu m^{-2}$, $V=80\,\rm V$).  In the clusters, the vast majority of the rods stand in the direction of the $\mathbf E$ field.
    Note that only 25\% of the rods are fluorescent, see Appendix A. 
    (e) Three subsequent pictures showing the capture and release of active rods at the interface between the arrested and active-liquid  phases.
    Motile rods may switch to the standing  state when colliding with the clusters. 
    Similarly, arrested rods can resume their self-propulsion and escape the cluster boundary. 
    (f) The area fraction of the arrested phase $\Phi_{A}$ grows when $\rho_0$ increases ($V=150\,\rm V$). Fitting the coexistence points with a straight line allows us to define the two binodals $\rho_{\rm L}$ ($\Phi_A=0)$ and $\rho_{\rm A}$ ($\Phi_A=1)$, corresponding to the vertical shaded lines.
    (g) At the onset of solidification all the $\Phi_{A}(\rho_0-\rho_{\rm L})$ collapse on a master curve (where the binodal $\rho_L$ is defined in {(f)}, yellow vertical line)
    (h) Phase diagram of interacting Quincke rods. The three macroscopic regions correspond to the homogeneous active liquid, the homogeneous arrested solid and to the coexistence region.
    }
    \label{Fig2}
\end{figure*}
We summarize the phase behavior of interacting Quincke rods in Fig.~\ref{Fig2}. 
At low packing fraction, the rods form an isotropic active liquid featuring no global orientational order, Fig.~\ref{Fig2}a, and no spontaneous macroscopic flow.
The lively yet uncorrelated dynamics of this active liquid is illustrated in Supplementary Movie 2.
One remarkable feature is worth being highlighted. 
Unlike isolated particles, when they interact, our Quincke rods do not remain horizontal and active throughout the entire experiment.
Instead, they  switch their self-propulsion on and off by changing their orientation and pointing in the direction of the $\mathbf E$ field. 
Using fluorescence microscopy, we easily localize the standing rods  that form very bright dots on our images, see Fig.~\ref{Fig2}b,~\ref{Fig2}d and ~\ref{Fig2}e.

Although the transition rates from the rolling to the standing state ($\alpha$) and from the standing to the rolling state ($\beta$)  cannot be measured individually, the fraction of standing rods is given by $\alpha/(\alpha+\beta)$ and  provides a  measure of their ratio, see Fig.~\ref{Fig2}c.
Despite the small statistics, we find that $\alpha$ is vanishingly small below a threshold density $\brho \simeq 0.1\,\rm\mu m^{-2}$, and increases  above this threshold. 
This behavior implies that the spontaneous transitions between the rolling and the arrested states is not a single-rod property. Instead, it results from the physical couplings between the Quincke rods that are mediated by contact, electrostatic and hydrodynamic interactions.
Despite the lack of gene regulation and chemical signaling,
the rod activity depend on the local number of neighboring particles.
This seemingly mundane observation is at the core of our work:  Quincke rods achieve a primitive form of quorum sensing.

Increasing $\rho_0$, or lowering the actuation voltage $V$, a second phase behavior emerges from these quorum sensing interactions. 
A finite fraction of rods switches to the arrested state, stand up and form dense solid clusters hosting no inner dynamics, Fig.~\ref{Fig2}d and Supplementary Movie 4.
These clusters of static standing  particles coexist with a homogeneous active liquid.
We observe a dynamical exchange of particles at their interface, see Fig.~\ref{Fig2}e and Supplementary Movie 5.
Increasing $\rho_0$ results in an increase of the area fraction of the  arrested phase $\Phi_{A}$ until  all the rods stop propelling and form a homogeneous arrested phase. 

To define more quantitatively the boundaries between the three different states, we plot   $\Phi_{A}$ as a function of $\rho_0$, Fig.~\ref{Fig2}f.
We perform a linear fit of  the $\Phi_A(\rho_0)$ curve in the coexistence region. 
The vertical lines  define the location of the phase boundaries, corresponding to $\Phi_A=0$ (Liquid) and $\Phi_A=1$ (Arrested). 
We also stress that the $\Phi_{A}(\rho_0)$ curves can all be collapsed onto a single master curve  at the onset of the solid-clusters nucleation, see Fig.~\ref{Fig2}g.
Repeating $84$ different experiments, we then build the phase diagram of the interacting Quincke rods shown in Fig.~\ref{Fig2}h.

\subsection{Quorum-sensing-induced phase separation}
\begin{figure*}
    \includegraphics[width=\textwidth]{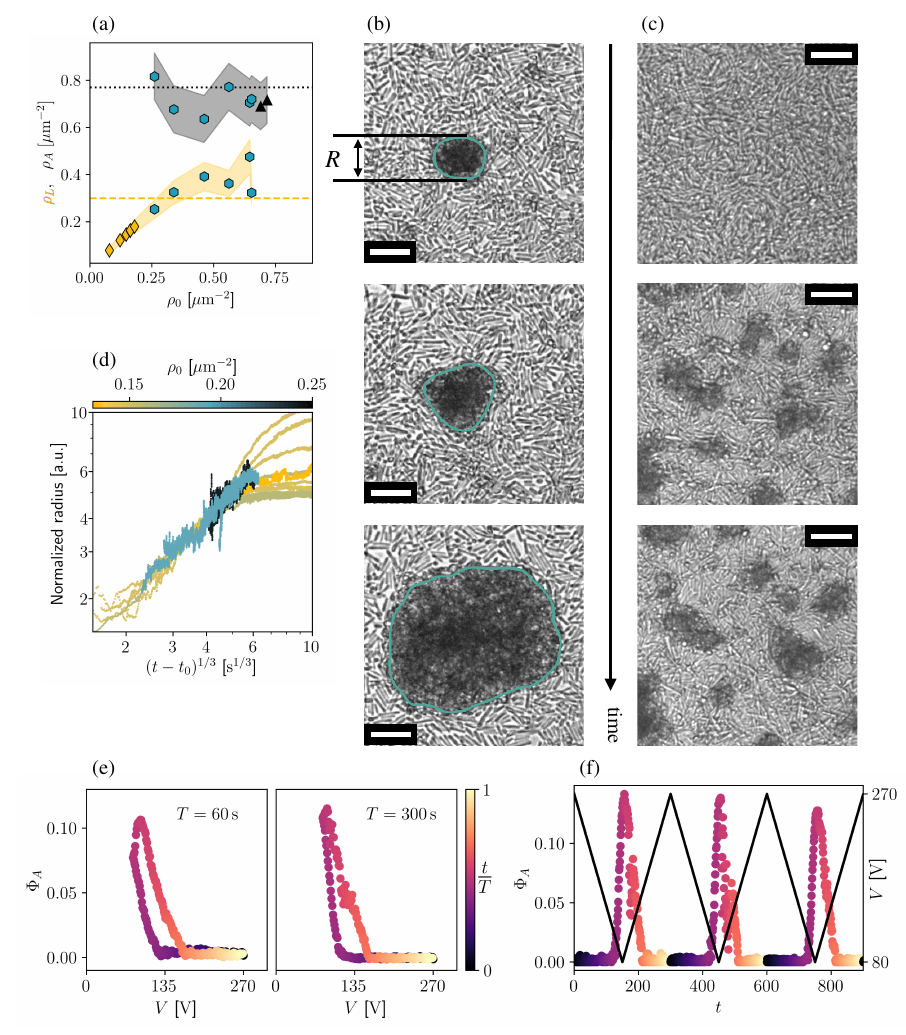}
    \caption{{\bf Quorum sensing induces phase separation}.
    (a) Density of the active liquid ($\rho_{\rm L}$) and of the arrested phases ($\rho_{\rm A}$) plotted against the  average rod density $\rho_0$. Horizontal lines correspond to the binodals determined in Fig.~\ref{Fig2}f. Voltage: $V=150\,\rm V$.
     (b) Growth of a a single arrested cluster in a homogeneous active liquid. 
     The typical size $R(t)$ is defined as the square root of the cluster area. 
     $V=109\,\rm V$, $\rho_0 = 0.15\,\rm \mu m^{-2}$.
     Scale bar: $5\,\rm \mu m$.
    (c) Growth kinetics showing the classical $R(t)\sim (t-t_0)^{1/3}$ scaling law. 
    (d) Image sequence showing the coarsening patterns upon a  quench deep in the coexistence region, from $V=300\,V$ to $V=90\,V$. $\rho_0=0.20\,\rm \mu m^{-2}$. 
    Scale bar: $5\,\rm \mu m$.
    (e) Hysteresis loops. 
    When the motorization voltage varies periodically the fraction of the arrested phase displays a hysteresis loop. 
    $\rho_0=0.2\,\rm \mu m^{-2}$.
    The two hysteresis cycles corresponds to different periods $T$ of the applied voltage. The voltage varies from $80\,\rm V$ to $270\,\rm V$.
    (Left):  $T=60\,\rm s$.
    (Right):  $T=300\,\rm s$.
    (f) Corresponding time series of $\Phi_{A}(t)$ and of $V(t)$. ($T=300\,s)$
    }
    \label{Fig3}
\end{figure*}
Our first set of observations hints towards a genuine phase separation scenario.
To further confirm this hypothesis, we run three sets of additional measurements: (i) we measure the local packing fraction within the two coexisting phases, (ii)  the phase ordering kinetics of the solid phase, and (iii) we cycle the magnitude of the $\mathbf E$ field to test the possibility of a hysteresis behavior. 
(i) In Fig.~\ref{Fig3}a, we plot the space and time averaged number density in the liquid and arrested phases ($\rho_{\rm L}$ and $\rho_{\rm A}$ respectively).
We find that they hardly depend on the average density $\rho_0$, as in  equilibrium when a solid phase coexists with a liquid. 
Even if the variations of $\rho_{\rm L}$ are noisy, they feature a clear slope change as the first solid clusters form, thereby signaling the coexistence between the two distinct phases.

(ii) To quantify the growth kinetics of the arrested 
phase we proceed as follows. 
We prepare the system deep in the liquid phase at different values of $\rho_0$ and set $V=275\,\rm V$. 
We then slowly reduce the value of the applied voltage below the coexistence line until a first solid cluster forms (Fig.~\ref{Fig3}b). 
We then keep the value of $V$ unchanged and measure the variations of the solid cluster area $R^2(t)$ over time, Fig.~\ref{Fig3}d. 
Repeating the same experiments $17$ times, we find that the growth kinetics is consistent with a $R(t)\sim t^{1/3}$ scaling law.
This scaling behavior is typical of a nucleation and growth dynamics  in equilibrium phase separations~\cite{Bray2002}. 
Repeating the same experiments, but lowering quickly the motorization voltage, we quench the system deep in the coexistence region, and observe a markedly different pattern, Fig.~\ref{Fig3}c. 
This second scenario  mirrors the morphology of the spinodal decomposition patterns observed in equilibrium and active systems~\cite{Bray2002,Stenhammar2013,vanderLinden2019,thompson2011lattice} as well as the formation of fruiting bodies in bacterial colonies~\cite{Liu2019}.

(iii) A robust test of phase metastability consists in studying the oscillatory response of a system across the boundary separating two distinct phases.
Here, we prepare the system at a density $\rho_0=0.18\,\rm \mu m^{-2}$ and apply a constant voltage $V=270\,\rm V$ until a steady  homogeneous liquid phase forms. 
We then apply periodic triangular modulations of $V$, of amplitude $\Delta V=190\rm V$, and measure the area fraction of the arrested phase.
We repeat this experiment  at various rates and plot in Fig.~\ref{Fig3}e  the instantaneous value of $\Phi_{A}(t)$ against $V(t)$.  
Fig.~\ref{Fig3}f show the corresponding time series.
We find an unambiguous hysteresis behavior over a range of modulation period ranging from 20~s to 300~s.
This observation provides a direct proof of metastability.

Altogether our three observations confirm our hypothesis: the phase behavior summarized in Fig~\ref{Fig2}h results from a genuine phase separation between and active liquid and an arrested  phase.

\section{Theory}
Building a mathematical model of Quincke rods that accurately captures contact, electrostatic, and hydrodynamic interactions remains beyond the reach of current theories and simulations.
To gain insight, we therefore construct a robust minimal model that isolates the essential ingredients needed to explain—and generalize—our experimental findings.
 
 We model our Quincke rods as 2D active Brownian particles (ABPs) that can switch between two states: a rolling state ($R$), where particles propel at constant speed $v_0$, and a standing state ($S$), where they are immobile. 
As above, we denote by $\alpha$ the rate at which particles switch from the rolling to the standing state, and by $\beta$ the rate at which standing particles resume their rolling motion in  random direction. 
Since the transition to the rolling state is induced only by collisions with rolling particles, we posit that $\beta$  depends only on the local density of rolling particles, which we denote by $\rho_R(\mathbf r)$ (in practice we compute it within a radius set by the rod length $\ell_r$). 
By contrast,
the transition from standing to rolling state is induced by any collision.
We therefore expect $\alpha$ to depend on the local density of particles $\rho(\mathbf r)$. 
In addition, Fig.~\ref{Fig2}b shows that $\alpha$ vanishes when $\rho<\bar \rho$. 

Informed by our observations and measurements, we hence model the transition rates as:
\begin{eqnarray}
  \label{eq:alpha-rods}
    R \xrightarrow{\alpha} S\;, &&\quad \alpha({\rho}) \>\>\>= \alpha_0 ({\rho}-\brho) \Theta({\rho}-\brho) \\[0.2cm]
    S \xrightarrow{\beta} R\;, &&\quad \beta({\rho}_R) = \beta_0 {\rho}_R \;,
  \label{eq:beta-rods}
\end{eqnarray}
where $\Theta$ is the Heaviside function.

\subsection{A Motility-Induced Absorbing Phase Transition}

\begin{figure*}
  \begin{center}
    \def\w{6.5cm}
    \includegraphics[width=\textwidth]{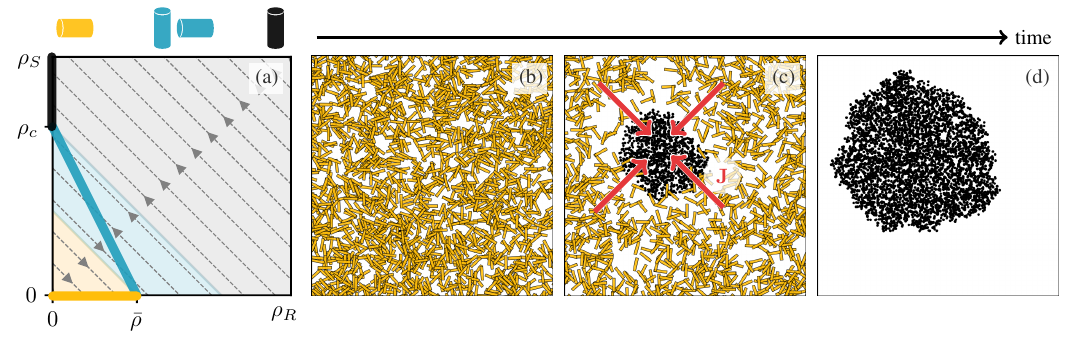}
    \caption{{\bf Quorum-sensing-induced absorbing phase transition}.
    (a) Streamplot of the mean-field dynamics of Eq~\eqref{eq:MF_rel}. The dashed gray lines correspond
      to the  dynamics, while solid lines represent the fixed points. 
      For $\rho_0<\brho$,  the active phase is the only stable attractor (yellow line, $\rho_S=0$). 
      At intermediate densities $\brho < \rho_0 < \rho_c$, the fixed points correspond to mixed configurations (teal line). At large densities, $\rho_0>\rho_c$, the static phase is the only stable attractor (black line).
      (b-d) Snapshots from a simulation of ABPs with density-dependent transition rates (Eqs.~\eqref{eq:alpha-rods} and~\eqref{eq:beta-rods}). 
      To facilitate comparison between experiments and simulations, we plot the rolling rods as golden rectangles and the standing rods as black circles. We show in SM that simulations of rods interacting via actual anisotropic quorum-sensing interactions lead to identical results.
      When an arrested cluster nucleates, it progressively absorbs the whole system as the particle current $\bfJ \propto - \nabla_{\bfr} \rho_R$.   
      In this panel, the length of the rods corresponds to the
      interaction radius $\ell_r=5$. 
      Numerical details: see Appendix~\ref{app:numerics} and SM.  \if{$\alpha_0=20$, $\beta_0=10$,
      $D_r=2$, $\rho_0=\bar{\rho}=0.20$. System size $150 \times 150$.}\fi }
    \label{fig:1TH}
  \end{center}
\end{figure*}
We first study the mean-field dynamics of our model to identify the stable homogeneous phases.
We denote by $\rho_0$ the homogeneous particle density. The corresponding equations for $\rho_R$ and $\rho_S$ are:
\begin{equation}
\dot \rho_R=\beta(\rho_R) \rho_S-\alpha(\rho_0) \rho_R\;\quad\text{and}\quad\dot \rho_S=-\dot \rho_R.
\label{eq:MF_rel}
\end{equation}
The second equation results from mass conservation and implies that the dynamics takes place along the line $\rho_S = \rho_0 -\rho_R$. 

First, we note that homogeneous arrested  phases with $\rho_R = 0$ are  linearly unstable when $\alpha_0 < \beta_0$ (See SM~\cite{supp}). 
This is incompatible with our experimental observations, we therefore focus solely on the regime $\alpha_0 > \beta_0$. 
In Fig.~\ref{fig:1TH}, we illustrate  the dynamics  of Eq.~\eqref{eq:MF_rel}, and the stable homogeneous phases it predicts. 
In agreement with our experiments, at small density ($\rho_0 < \bar{\rho}$) the  stable attractors define an active phase where all the particles roll and $\rho_S=0$ (yellow line).
Similarly, at large densities ($\rho_0>\rho_c\equiv \alpha_0\bar{\rho}/(\alpha_0-\beta_0)$), we predict a fully-static phase where $\rho_R=0$  (black line).
However, in the intermediate density range ($\bar\rho<\rho<\rho_c$) the mean field dynamics lead to a stable homogeneous mixture of standing and rolling rods, which we never observe in our experiments. 

To account for phase separation, we now go beyond the mean-field approximation and
consider the impact of spatial fluctuations.
As detailed in SM, we build the large scale dynamics of ABPs that can switch their activity on and off in response to quorum-sensing interactions.
The switching rates are given by Eqs.~\eqref{eq:alpha-rods} and~\eqref{eq:beta-rods}, and
the resulting dynamics reduces to a diffusion equation  driven by the density of rolling particles:
\begin{equation}\label{eq:diff dyn}
    \partial_t \rho(\bfr,t) = -\nabla_\bfr \cdot \bfJ\;,\qquad \bfJ=-\cD_{\rm eff} \nabla  \rho_R\;,
\end{equation}
where $\bfJ$ denotes the total  current and $\cD_{\rm eff}$ is the large scale diffusivity of the rolling particles. 
This minimal model is however not enough to account for the observed coexistence between an active liquid (where $\rho_R\neq0$) and a fully arrested phase (where $\rho_R=0$).  
 The flux-free condition indeed requires
$\rho_R$ to be homogeneous, and the dynamics results in  an absorbing phase transition illustrated by numerical simulations in Fig.~\ref{fig:1TH}b-d and Supplementary Movie 6.
When a cluster of standing particles nucleates in an active liquid, Eq.~\eqref{eq:diff dyn} implies that a net particle flux drives the rolling particles into the  cluster, thereby leading an to unstoppable growth of the arrested phase (Fig.~\ref{fig:1TH}c).

We have learned that quorum sensing alone cannot explain our experimental observations. 
In experiments, however, the  current is not only determined by the active displacement of the particles but also by their steric repulsion.
We show below that these  repulsive forces compete with quorum sensing and play a pivotal role in stabilizing phase coexistence.

\subsection{Steric repulsion  hinders  motility-induced absorbing phase transition}
\begin{figure*}
  \begin{center}
    \def\w{6.5cm} \includegraphics[width=0.9\textwidth]{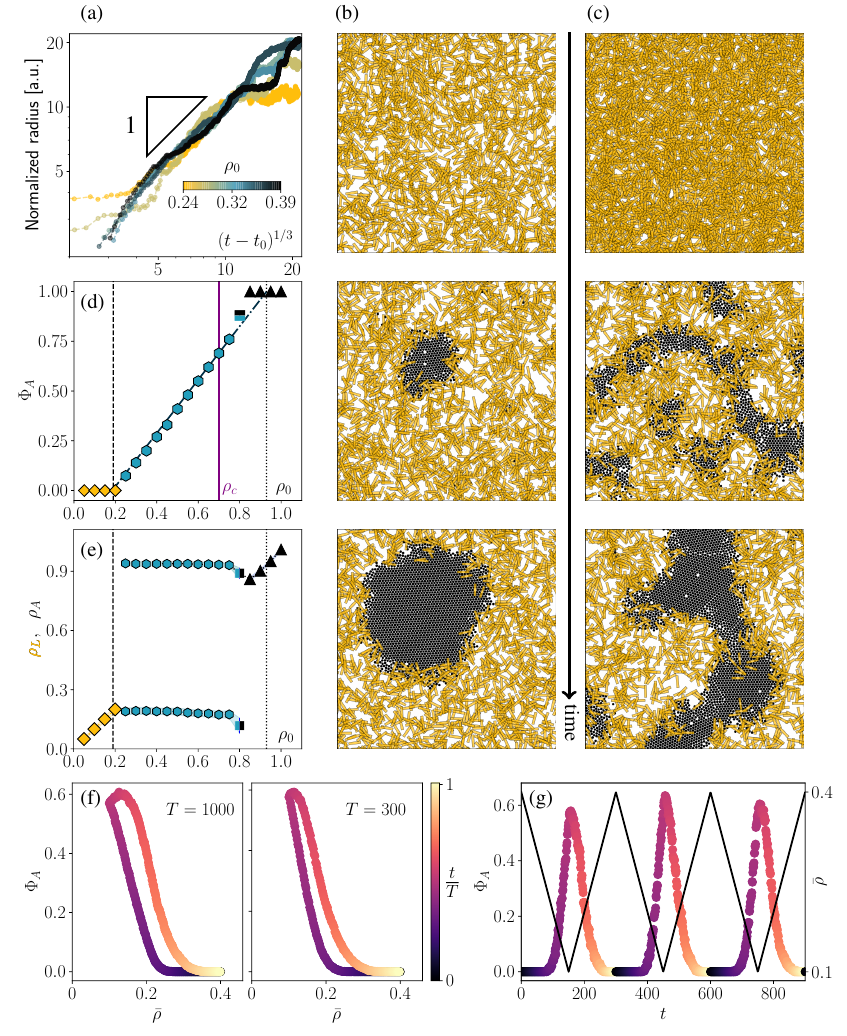}
    \caption{{\bf  Competition between quorum sensing and pairwise repulsion in numerical simulations}. (a) Coarsening of static domains. (b) Nucleation of a static bubble at $\rho_0=0.24$. (c) Spinodal decomposition at $\rho_0=0.45$. (d) Static area fraction $\Phi_{A}$ as
      a function of $\rho_0$. The solid vertical line corresponds to the
      critical density $\rho_c$ beyond which the mean-field
      dynamics predicts a rapid convergence to a homogeneous
      static phase. The two dashed vertical lines represent the density of coexisting phases. Yellow diamonds correspond to a fully active gas; teal hexagons to phase coexistence; teal and black squares represent metastable regions where both the arresting phase transition and phase coexistence are observed; black triangles to a static solid.  (e) Liquid and solid binodals measured in simulations. We use the same convention for symbols as in panel d. (f) Hysteresis loop for the static area fraction as a function of $\bar{\rho}$,
      \textit{i.e.} the threshold beyond which the standing rate
      $\alpha(\rho)>0$. In simulations, the activity of the system is varied in time via a sawtooth protocol for $\bar{\rho}(t)$ and $v_0(t)$ with period $T$ as shown in the black curve of panel (g). (g) Time evolution of $\Phi_{A}$ under a sawtooth protocol for $\bar{\rho}(t)$ and $v_0(t)$. Numerical details: see Appendix~\ref{app:numerics} and SM.}
    \label{fig:2TH}
  \end{center}
\end{figure*}
Keeping in mind that we are looking to build a minimal theory, 
we model the repulsive forces between the rods with a  
Weeks-Chandler-Andersen interaction potential
 $     U(r) = \big\{4 E_0 r^{-12} [{\sigma^{12}} - (\sigma r)^6] +E_0\big\} \Theta(r_0 - r)$,
where $\sigma$ represents the particle size, $r_0 = 2^{1/6} \sigma$.
The force that particle $i$ exerts on $j$ is denoted as
$\bff_{ij}$. 
For simplicity, the interactions are here isotropic and independent of the particle state ($R$ or $S$). 
In our experiments the standing rods are densely packed in the static phase,  we hence take $\sigma<\ell_r$. 
In the rolling state, the position $\bfr_i$ and  orientation $\bu_i=(\cos\theta_i,\sin\theta_i)$ of particle $i$  evolve according to
\begin{eqnarray}
  \label{eq:R-dyn-PFAP}
\dot{\bfr}_{i} = v_0 \bfu_i(\theta_i) - \mu \sum_{j\neq i} \nabla U(\bfr_i-\bfr_j),\,\quad \dot{\theta}_i = \sqrt{2 D_r} \eta_i\,,
\end{eqnarray}
where $\mu$ is the particle mobility and $\eta_i(t)$ is a centered Gaussian white noise with unit variance. 
The static particles also evolve according to~Eq.~\eqref{eq:R-dyn-PFAP}, albeit with $v_0=0$. 

The phenomenology revealed by simulations of Eq.~\eqref{eq:R-dyn-PFAP} is  qualitatively similar to our experimental observations: the interplay between steric repulsion and quorum sensing (Eqs.~\eqref{eq:alpha-rods}-\eqref{eq:beta-rods}) arrest the absorbing transition described in Fig.~\ref{fig:1TH}, and  lead to the  coexistence between an active liquid  and a dense arrested phase (see Fig.~\ref{fig:2TH}b,c and Supplementary Movies 7---8). 
We stress  that the similarities between simulations and experiment are not restricted to the steady states: 
our model also quantitatively reproduces the scale-free coarsening dynamics $R(t) \sim t^{1/3}$ (Fig.~\ref{fig:2TH}a).

To further confirm the relevance of our model, in Fig.~\ref{fig:2TH} we compute  the evolution of $\Phi_A$ with $\rho_0$,  and confirm the existence of hysteresis  upon periodic drive. 
We first vary the average density $\rho_0$ and report in Figs.~\ref{fig:2TH}d the  area fraction of the arrested phases  $\Phi_{ A}(\rho_0)$ as well as the coexisting densities in Figs.~\ref{fig:2TH}e. 
As in a standard liquid-gas phase coexistence, the density of the active-liquid ($\rho_L$) and of the arrested phase ($\rho_A$) remain constant as $\rho_0$  varies.
As a result, the lever rule is satisfied in Fig~\ref{fig:2TH}d.
These predictions are consistent with our measurements reported in Fig.~\ref{Fig3}.
Finally, to mimic the effect of varying the electric potential $V$, we modulate both $v_0(V)$ and $\brho(V)$ in time using the sawtooth protocol illustrated in Fig.~\ref{fig:2TH}f. 
The time evolution of the static area fraction $\Phi_A(t)$ then reproduces the hysteresis loops observed in experiments (Fig.~\ref{fig:2TH}e and Supplementary Movie 9). 
We stress the robustness of our findings  in SM. We show that all the phenomena reported in Figures 3 and 5 are not specific to monodisperse  particles interacting isotopically.

Our minimal simulations demonstrate that the competition between quorum sensing and repulsive forces leads to an unanticipated phase-separation mechanism in scalar active matter. 
It is markedly different from the well-known Motility-Induced Phase Separation (MIPS)~\cite{cates2015MIPS}:
Firstly, repulsive forces act against the proliferation of the low motility phase.
Secondly, as illustrated by the hysteresis loops, increasing activity favors homogeneous systems instead of inducing phase separation.   
Finally,  as $\rho_0$ exceeds $\rho_c$, contact interactions are unable to counter the effect of quorum sensing.
Fig.~\ref{fig:2TH}d shows that phase coexistence (teal symbols) is rapidly replaced by an absorbing arrested state (black symbols) not captured by MIPS physics.
 Supplementary Movies 10---11 shows that close to $\rho_c$ both phases are metastable. 
 Depending on the initial conditions, the system may either remain fluid and freeze (mixed symbols in  Fig.~\ref{fig:2TH}d and~\ref{fig:2TH}e).
While we do not control the initial condition of our experiments, we note that Fig.~\ref{Fig2}g features a  sharp transition to the arrested phase similar to our simulations.  
This behavior suggests that, at high densities, the  coexistence between rolling and standing particles is ultimately suppressed  in favor of an absorbing phase transition. 

In the next section we develop a theory that explains how the competition between quorum sensing and repulsive interactions can shape the phase behavior of  active matter.

\subsection{Explaining phase coexistence through generalized thermodynamics}
From now on, we focus on the coexistence regime and show how it can be accounted for by a generalized thermodynamic theory. 
To gain more insight we first discuss how quorum sensing and pairwise forces control the particle flows.  
In SM, we show that pairwise interactions modify the  current given by Eq.~\ref{eq:diff dyn} into
\begin{equation}
    \bfJ = \mu \nabla_{\bfr} \cdot  \bfsigma_a+ \mu \mathbf{\Delta}+\mu \nabla_{\bfr} \cdot \bfsigma_{\rm IK}\;. 
  \label{eq:generalized-stress1}
\end{equation}
As in systems of ABPs interacting via pairwise forces, $\bfsigma_a$ is a so-called  active stress~\cite{solon2015pressure,speck2021coexistence}. 
In addition, quorum-sensing interactions induce another active flux  $\mathbf{\Delta}(\bfr)\propto \nabla \alpha[\rho(\bfr)]$ that cannot be expressed as the divergence of a local stress. 
As in Eq.~\eqref{eq:diff dyn}, the sum of these two active fluxes scales as $-\nabla\rho_R$ and therefore
drives the particles from the active-liquid phase to the arrested clusters, see SM.
However, in the presence of repulsive forces between the particles, the total flux is enriched by $\nabla_{\mathbf r}\cdot\bfsigma_{\rm IK}$, where $\bfsigma_{\rm IK}$ is the Irving-Kirkwood stress tensor that stems from the pairwise conservative interactions~\cite{irving1950statistical}.
This stress results in a flux from the dense arrested phase towards the dilute active liquid that opposes the active flux.
This competition qualitatively explains the arrest of the absorbing phase transition, and the stabilization of  the interfaces between the two phases.

This intuitive picture can be turned into a quantitative theory by performing a gradient expansion of the hydrodynamic equations~\eqref{eq:generalized-stress1} (see SM).
To leading order, we can express the dynamics for $\rho$ as a gradient descent in an effective free-energy landscape $\cG(\bfr, [\rho])$:
\begin{equation}\label{eq:thermo}
    \dot{\rho} = \nabla_{\bfr} \cdot \left[ \mu \rho \nabla_{\bfr} \frac{\delta \cG}{\delta \rho(\bfr)}\right] \;.
\end{equation}
As in equilibrium, phase separation occurs when $\cG$ features at least two minima (see inset of Fig.~\ref{fig:press}a).
We can then use a common-tangent construction to predict the values of the coexisting densities $\rho_L, \rho_A$~\cite{cates2015MIPS,solon_2018_generalized}. 
In Fig.~\ref{fig:press}d, we compare our numerical and theoretical predictions.
The good agreement confirms  that our effective equilibrium theory correctly predicts how the densities of  the active liquid and arrested phases  vary with the threshold density $\brho$. 

\subsection{Mechanical equilibrium at unequal pressure}
\label{sec:mechanics}
\begin{figure*}
    \includegraphics[width=\textwidth]{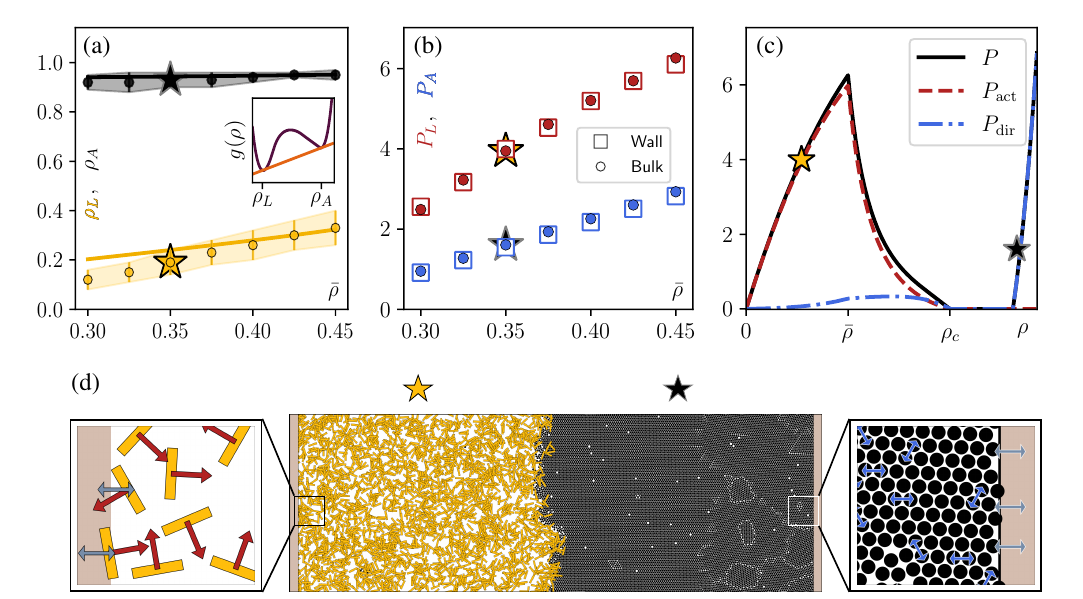}
    \caption{{\bf Phase coexistence and pressure imbalance across flat interfaces}. (a) Coexisting densities for the liquid ($\rho_L$) and arrested ($\rho_A$) phases: comparison between local theory (solid lines) and numerical simulations (points) as a function of $\brho$. Star symbols correspond to the coexisting densities for the simulation of panel d. In the inset we report the common-tangent construction on the generalized free-energy density $g(\rho)$ at $\brho=0.35$. (b) Comparison between bulk and wall pressure measured in phase-separated profiles, plotted as a function of the threshold density $\bar{\rho}$. The bulk pressure of the active liquid, $P_L$, and of the arrested phase, $P_A$, are equal to the mechanical pressure on the neighboring wall. At phase separation, the two bulk pressures are not balanced and $P_L > P_A$. (c) Theoretical predictions for the bulk pressure $P = P_{\rm act} + P_{\rm dir}$ as a function of the density $\rho$. The bulk pressures $P_L$ and $P_A$ measured in simulations are reported with star symbols, falling on top of our theoretical curve. (d) Phase-separated configuration of motile rods (yellow rectangles) and static particles (black circles). In the motile liquid phase, $P_L$ is dominated by the active contribution $P_{\rm act}$, as sketched on the left inset. In the arrested phase, the bulk pressure $P_A$ is given by a direct-pressure term $P_{\rm dir}$, resulting from pairwise repulsion (see right inset). Numerical details: see Appendix~\ref{app:numerics} and SM.}
    \label{fig:press}
\end{figure*}
To close our article, we  discuss more broadly the atypical mechanics and thermodynamics that emerge from the competition between quorum-sensing and contact interactions. 
We provide all technical details in SM.

To explore the mechanical properties of the interface that separates the active liquid and arrested phases, we consider the basic setting sketched in Fig.~\ref{fig:press}d: 
a flat interface separates two homogeneous phases in contact with  confining walls.  
We now show that both phases apply a finite pressure on the confining walls that corresponds to their bulk pressure, as in equilibrium.
However, the flux-free state where the interface is immobile corresponds to a situation where a net pressure drop forms, even when the interface is flat.
To understand  this nonequilibrium property, we note that in the active-liquid phase $\rho<\bar\rho$.
Therefore, the transition rate from the rolling to the standing state vanishes ($\alpha=0$), and so does the active flux:  $\Delta =0$. 
As for standard Active Brownian Particles, the pressure exerted by the active liquid on the left confining wall ($P_L$) thus satisfies an equation of state given by the sum of a direct contribution, $P_{\rm dir} \equiv -\sigma_{\text{IK},xx}$, and of the active pressure, $P_{\rm act} \equiv -\sigma_{a,xx}$. 
In the arrested phase, $\rho_R=0$ and the pressure satisfies an even simpler equation of state: $P_A=P_{\rm dir}$. 
These bulk pressures are equal to the force density exerted on the confining walls.
Fig.~\ref{fig:press}b indeed shows the perfect agreement between their numerical measurements and therefore confirms the existence of an equation of state that characterizes each phase separately. 
In SM we determine analytically closed expressions for the active and direct pressures~\cite{speck2021coexistence,martin2021statistical}, which are shown in Fig.~\ref{fig:press}c. 
We again find an excellent agreement with our numerical measurements at coexistence.

However, at the interface between the two phases, $\mathbf{\Delta}$ does not vanish.
As a consequence, the bulk pressures in the active-liquid and arrested phases must be different. 
To see this, we project Eq.~\eqref{eq:generalized-stress1} along $x$ and integrate it across the interface to compute the pressure drop:
\begin{eqnarray}
    P_L - P_A = \int_{x_L}^{x_A} \Delta_x(x) \rmd x \;, 
    \label{eq:P_balance}
\end{eqnarray}
where $x_{L,A}$ are abscissa deep in the liquid and arrested phases. 
Keeping in mind that $\Delta_x(x) \propto\partial_x\alpha(\rho)$, 
 Eq.~\eqref{eq:P_balance} readily tells us that the pressure drop originates from the quorum-sensing interactions localized at the interface.
This counterintuitive results has a clear microscopic explanation.
In the bulk of the active-liquid phase,   the switching rate is vanishingly small and the particles keep on rolling. 
Conversely, as they approach the dense arrested phase, they stochastically arrest and resume their self-propulsion.
This switching process results in a reduction of the net forces they apply on the standing particles at the interface.
The pressure at the interface is thus lower than in the liquid bulk:  $P_L>P_A$.
Simply put, a finite pressure drop can build across flat, but active, interfaces. 

\section{Conclusion and outlook}  
We have demonstrated that colloidal rods powered through Quincke electrorotation show  primary forms of  sensing and adaptation to crowding. 
Studying their phase behavior, we have shown that the active rods  self-organize  into phase-separation patterns typical of living bacterial colonies~\cite{Liu2019}.
We have theoretically explained this robust phase separation phenomenon based on the competition between quorum sensing and contact interactions:
Quorum sensing generically drives interacting active particles towards an absorbing state where they collectively lose their motility. 
However, this absorbing phase transition can be  hindered by physical contacts. 
The outcome of this competition yields an unanticipated form  of macroscopic phase-separation where fully active populations coexist with macroscopic groups of particle that spontaneously stop their inner motorization. 
We believe that our work is  a first step towards the realization and modelling of adaptable active matter where activity and self-organization are two-way coupled.

\section*{Acknowledgments}
We thank Alexandre Morin for illuminating discussions. This work was partly supported by the European Research Council (ERC) under the European Union’s Horizon 2020 research and innovation program (grant agreement No. [101019141]) (DB). 
AD acknowledges an international fellowship from Idex Université Paris Cité and a postdoctoral contract from the University of Geneva.

\section*{Data availability}
The data that support the findings of this article are openly available~\cite{lefranc_2025_15630142}.

\appendix

\section{Quincke roller experiments}\label{app:Quincke}
We first follow the protocol introduced in~\cite{Fernandez2019} to synthesize 
$3.7\pm1.5\,\rm \mu m$ long photoresist cylinders (SU-8 50) of diameter $0.55\pm0.17\,\rm \mu m$ (aspect ratio: 6.7).
We dye 25\% of the  colloids with  Rhodamine 110 (Sigma) and use OLOA 1100 stabiliser (4wt\%) to transfer them  from their aqueous solvent to a mixture of Hexadecane oil and Dioctyl sulfosuccinate sodium salt ($0.12\,\rm mol/l$).

We inject the rods in a tilted  V-shape microfluidic channel and let them sediment for six hours,  as sketched in Fig.~\ref{Fig1}a. 
The channel height and width  are $25\,\rm \mu m$ and $1\,\rm mm$ respectively. 
We then place our device on a microscope stage and  observe that the colloidal rods accumulate at the bottom surface over one to two  layers.  
To make the channels we cut a $25\,\rm \mu m$-thick adhesive films (Teraoka) with commercial plotting cutter (Graphtec). 
The top and bottom walls are  two transparent electrodes made of  glass slides coated with Indium Tin Oxide (Solems, ITOSOL30; thickness, 80 nm). 

As expected, we  observe variations of  the local number density  $\rho(\mathbf r)$ along the channel. 
We measure  $\rho(\mathbf r)$ manually using confocal fluorescence microscopy. 
It is defined as $\rho(\mathbf r)=N(\mathbf r)/A$, where  $N(\mathbf r)$ is the number of colloids in a  voxel of height equal to the channel height, and of base area $A=150\,\rm \mu m^2$.  

Applying a DC voltage $V$  across the two electrodes that form the top and bottom walls, we then turn our plastic rods into motile bodies powered by the Quincke electrohydrodynamic instability, see Supplementary Movies 1 and 2.
We apply a stable voltage ranging from 10 V to 300 V with Trek 609E-6 amplifier.

We observe the dynamics of the active rods with an inverted microscope (Zeiss Axio Observer A1) and a 40X objective (NA$=0.65$).
We take both bright field and fluorescence images with a Ximea MC124MG-SY-UB camera (frame rates 100, or 20 fps).
To measure the local number density of the rods we use a confocal microscope (Thorlabs) with MCLS laser source (482 nm) and a 525 nm emission filter (pinhole diameter $1.2\,\rm mu m$). 

\section{Numerical Methods}\label{app:numerics}
We describe the algorithm used to perform particle-based simulations in a $2d$ continuous space and provide details on our numerical methods.

\subsection{Simulations in the absence of pairwise forces}
\label{sec:SQS-simuls}
We initialize our simulations in a homogeneous configuration with all particles in the rolling state, and  with random initial orientations. 
At each timestep $\rmd t$, when a particle is in the rolling (R) state, its position $\bfr_i$ and orientation $\bfu_i=(\cos \theta_i, \sin \theta_i)$ are updated according to: $\bfr_i(t+\rmd t) = \bfr_i(t) + v_0 \bfu_i(\theta_i) \rmd t$, $\theta_i(t+\rmd t) = \theta_i(t) + \sqrt{2 D_r \rmd t} \eta_i$, where $\eta_i$ is a Gaussian random number with zero mean and unit variance.
Particles in the standing (S) state do not change their position, and have no orientation.

To simulate the transitions between states, we proceed as follows. For each particle $i$, we first measure the local densities of standing and rolling particles. To do so, we use spatial hashing and divide the simulation domain in square boxes of linear size $\ell_r=5$. This allows us to measure the local density around each particle efficiently by looking solely at the box it belongs to and to the 8 neighboring boxes. The local densities of rolling (R) and standing (S) particles around particle $i$ are computed as:
\begin{eqnarray*}
    \rho_{R,i} = \sum_{j: \> r_{ij}<\ell_r}^{(R)} K(r_{ij}) \;, \quad \rho_{S,i} = \sum_{j: \> r_{ij}<\ell_r}^{(S)} K(r_{ij})\;,  \notag\\
    \text{where} \quad K(r) = \frac{1}{Z} \exp\biggl(-\frac{\ell_r^2}{\ell_r^2-r^2}\biggr) \Theta(\ell_r-r)
\end{eqnarray*}
is a normalized, smooth isotropic kernel with compact support on a disc of radius $\ell_r$. 
The sums $\sum^{(R)}$, $\sum^{(S)}$ are carried out over rolling and standing particles, respectively, and $r_{ij}$ denotes the distance between particles $i$ and $j$. Finally, the total density is defined as $\rho_i = \rho_{R,i}+\rho_{S,i}$. 
We then use Eqs.~\eqref{eq:alpha-rods}--\eqref{eq:beta-rods} of the main text to compute the standing rate $\alpha_i(\rho_i)$ and rolling rate $\beta_i(\rho_{R,i})$. 

We then determine whether a transition between states occurs once the particles have moved. For each particle, we draw a random number $q_i$ from a uniform distribution in $[0,1]$. If particle $i$ is in the rolling state and $q_i<\alpha_i \rmd t$, the particle enters the standing state. Similarly, a standing particle switches into the rolling state if $q_i<\beta_i \rmd t$.

\subsection{Simulations in the presence of pairwise forces}
\label{sec:SQS-PFAP-simuls}
At each timestep, we first compute the total force $\mathbf{f}_i$ acting on particle $i$. 
Since our WCA interactions are short-ranged, we use the same spatial hashing as for the density measurements, and take the interaction radius $r_0$ for WCA to be smaller than $\ell_r$. 
The position $\bfr_i$ of rolling particles is then updated as $\bfr_i(t+\rmd t) = \bfr_i(t) + \left[ v_0 \bfu_i(\theta_i) + \bff_i \right] \rmd t$, while for S particles $\bfr_i(t+\rmd t) = \bfr_i(t)  + \bff_i \rmd t$. 
The updates of the orientations of the rolling particles and the switching dynamics are identical as in the case where no pairwise forces act on the particles.

To prevent divergences due to the WCA potential, we use an adaptive timestepping method: we compute the maximal force acting on a particle, denoted as $f_{\rm max}$, and ensure that no particle moves by more than $\sigma/5$ within a time-step $\rmd t$. 
To do so, before updating the particle positions, we check whether $(f_{\rm max} + v_0) \rmd t < \sigma/5$; if not, we set $\rmd t = \sigma/[5 (f_{\rm max} + v_0)]$ for the current timestep.

\end{document}